\documentstyle[12pt]{article}

\begin{document}
\setcounter{page}{1}

\begin{flushright}
IPM-98-283 \\
gr-qc/9806007
\end{flushright}

\pagestyle{plain}
\vspace{1cm}

\begin{center} 
\Large{\bf Application of Colombeau's Generalized Functions to Cosmological              
Models with Signature Change}\footnote{contribution to the conference
ISMC 98, March 29-April 5, Potsdam, Germany.}\\
\small
\vspace{2cm}
Reza Mansouri\footnote{e-mail: mansouri@netware 2.ipm.ac.ir} and
Kourosh Nozari\footnote{e-mail: nozari@netware2.ipm.ac.ir}

{\it Institute  for Studies in Theoretical Physics and Mathematics, P.O.Box 
19395-5531 Tehran, Iran, \\
and  \\
Department of Physics, Sharif University of Technology, P.O.Box
11365-9161, Tehran, Iran.}\\
\today                         
\end{center}
\vspace{1.5cm}
\begin{abstract}
Colombeau's generalized functions are used to adapt the distributional
approach to singular hypersurfaces in general relativity with signature 
change. Equations governing the dynamics of singular hypersurface is obtained 
and it is shown that the stress-energy tensor of the surface can be non-vanishing.
\end{abstract}
\newpage
\vspace{1.5cm}
\section{Introduction}

Hartle and Hawking[1], in constructing a satisfactory model of the universe,  
try to avoid the initial spacetime singularity predicted by the standard model 
of cosmology using a combination of general theory of relativity 
and quantum mechanics. The basic features 
of the so called `` Hawking Universe'' obtained are as follows:
\begin{enumerate}
\item [1.]  A satisfactory theory of quantum gravity will represent the 
gravitational field, in the manner of general theory of relativity, by 
a curved spacetime[1].
\item [2.]  The proper understanding of ordinary quantum mechanics is provided 
by Feynman's ``path-integral'' or ``sum-over-histories'' 
interpretation. In ordinary quantum mechanics the basic 
idea is that a quantum particle does not follow a single ``path'' between two 
spacetime points, and so does not have a single ``history'', but rather we 
must consider all possible ``paths'' connecting these points. Therefore the 
usual wave function $\Psi$ is interpreted as an integral over all 
possible ``paths'' that a quantum system may take between two state. To solve 
the path integral, however, one must rotate the time variable in the usual 
quantum mechanical wave function to imaginary values in the complex plane, 
which yields as the new time coordinate the ``Euclidean'' time $\tau = it$ 
[1,2].
\item [3.]  There is a wave function for the entire universe $\Psi_{U} $ that 
is given by a Feynman path integral. The basic idea here is that one sums over 
all possible four-dimensional spacetimes( or spacetime ``histories'')  
connecting two three-dimensional spaces (states). In order to evaluate the 
path integral, however, one must again rotate the time variable to imaginary 
values which changes the integral from Lorentzian to Euclidean one.
The result is that the temporal variable in the wave function is changed to a  
spatial one. In other words, $\Psi_{U}$ sums only over Euclidean spacetimes, 
that is, over four-dimensional spaces with positive definite  
signature (+ + + + )[1-3].
\item [4.]  One wants to reach a certain state $S$ at which the evolution of 
the universe becomes classical, in accordance with general theory of relativity  
and standard model of cosmology. Accordingly, Hawking proposes a path integral 
over the Euclidean four-space $g_{\mu\nu}$, and matter-field configurations  
$\phi$ that yields $S$. $S$ is characterized by the three-metric $h_{ij}$ 
and a value of the scalar field, $\phi$.
\item [5.] To avoid an initial spacetime singularity, the cosmic path integral 
will include only compact(or closed) four-geometries, so that the 
three-geometries, marking successive states of the universe, shrink 
to zero in a smooth, regular way[6]. Hawking's universal wave function 
is obtained, therefore, by integrating only over compact four-geometries 
(Euclidean ``spacetimes'') that have the 3-space $S$ as the only (lower) 
boundary and are such that a universe in state $S$ will subsequently evolve.
\end{enumerate}
Statements (3) and (5) above are the essence of the idea that the 
universe was initially Euclidean and then, by change of signature 
of ``spacetime'' metric, the transition to usual Lorentzian spacetime 
occured. Earlier attempts to describe this interesting aspect was based on
Euclidean path integral formulation of quantum gravity and the analogy to
quantum tunelling effect in quantum mechanics[1]-[6].\\
The rise of this idea led many authors to consider it within the classical
theory of general relativity[7-20], with some controversies regarding 
the nature of the energy-momentum tensor of the hypersurface of signature
change[15,18,19]. Ellis and his coworkers[8], have shown that classical Einstein field 
equations, suitably interpreted, allow a change of signature of spacetime. 
They have also constructed specific examples of such changes in the case of 
Rabertson-Walker geometries. Ellis[9], by constructing a covariant 
formalism for signature changing manifolds, has shown the continuity 
of geodesics over the changing surface. Hayward[10] gives the junction 
conditions necessary to match a region of Lorentzian-signature spacetime 
to a region of Riemannian-signature space across a space-like 
surface according to the vaccum Einstein or Einstein-Klein-Gordon equations.
As we will show, Hayward's junction condition $\dot{R} = 0$ is not 
compulsory. He has also considered the increasing entropy, large-scale 
isotropy, and approximate flatness of the universe in the context of 
signature change[11]. Dereli and Tucker[12], described classical models 
of gravitation interacting with scalar fields having signature changing 
solutions. Kossowski and Kriele[13] considered smooth and discontinuous 
signature e e ange andnderived n necessaryry and sufufficient t junction c conditionons 
for r both proroposals. . They invesesgate thehextent totohich thehese are  
equivalalt. They y have clalaimed thahat non-flatat vaccum m sosolutions of 
Einstein equations can only occur in the case of smooth signature change.
Hellaby and Dray[15, 19] argue for a non-vanishing energy-momentum tensor 
of the signature changing surface. Hayavour a vanishing energy momentum tensor[18]. 
Kriele and Martin[16] do not accept the usual blief that signature change   
could be used to avoid space-time  singularities, except one abandon  
the Einstein equations at the signature changing surface.   
Martin[17] has studied Hamiltonian quantization of general relativity 
with the change of signature. He has also studied cosmological 
perturbations on a manifold addmiting signature change.\\
We intend to use the distributional method of Mansouri and Khorrami [21,23]
to approach the signature changing problem within the general theory of 
relativity. This is a powerfull formalism which can also be adapted to 
our problem, using the Colombeau's generalized theory of distributions[24-28].
This remedies the difficulties of non-linear operations of distributions 
in the framework of classical theory of Schwartz\_Sobolev[29]. In section
2 we give an overview of the Colombeau's generalized theory of distributions.
Sectoin 3 contains the adaptation of the distributional formalism for
singular surfaces in general relativity to the case of signature change,
followed by concluding remarks in section 4. \\
$Conventions$ $and$ $definitions$: We use the signature $(- + + + )$ 
for Lorentzian region and follow the curvature conventions of Misner, 
Thorne, and Wheeler (MTW). Square brackets, like $[F]$, are used to indicate 
the jump of any quantity $F$ at the signature changing hypersurface. As 
we are going to work with distributional valued tensors, there may be 
terms in a tensor quantity $F$ proportional to some 
$\delta$-function. These terms are denoted by $\hat{F}$.\\

\section{A short review of Colombeau theory}
Classical theory of distributions, based on Schwartz-Sobolev theory of
distributions, doesn't allow non-linear operations of distributions 
[29]. In Colombeau theory a mathematically consistent way of 
multiplying distributions is proposed. Colombeau's motivation  
is the inconsistency in multiplication and differentiation of
distributions. Take, as it is given in the classical theory of distributions, 
\begin{equation}
\label{math:2.1}
\theta^n = \theta \quad \quad \forall \ n = 2,3,\dots ,
\end{equation}
where $\,\theta\,$ is the Heaviside step function. Differentation of 
(\ref{math:2.1}) gives,
\begin{equation}
\label{math:2.2}
n \theta^{n-1} \ \theta' = \theta'.
\end{equation}
Taking $n=2$ we obtain  
\begin{equation}
\label{math:2.3}
2 \theta \theta' = \theta'.
\end{equation}
Multiplication by $\,\theta\,$ gives,
\begin{equation}
\label{math:2.4}
2 \theta^2 \theta' = \theta \theta'.
\end{equation}
Using (2) it follows
\begin{equation}
\label{math:2.5}
\frac{2}{3} \theta' = \frac{1}{2} \theta',
\end{equation}
which is unacceptable because of $\theta' \not= 0\,$. The trouble 
arises at the origin being the  unique singular point of $\,\theta\,$ and
$\,\theta'\,$. If one accepts to consider $\,\theta^n\not = 
\theta $ for $n = 2, 3, \dots$, the inconsistency can be removed.
The difference $\,{\theta}^n - \theta\,$, being infinitesimal, 
is the essence of Colombeau theory of generalized functions. 
Colombeau considers $\theta(t)$ as a function with ``microscopic structure"
at $t=0$ making $\theta$ not to be a sharp step function, but having 
a nonvanishing width, $\tau$[24]. $\theta(t)$ can cross the normal axis at any 
value $\epsilon$ where we have chosen it to be $\epsilon > \frac{1}{2}$[29]. 
It is interesting to note that the behaviour of $\theta(t)^{n}$ 
around $t=0$ is not the same as $\theta(t)$, i.e. $\theta(t)^{n} \not 
= \theta(t)$ around $t=0$[24]. In the following we give a short 
formulation of Colombeau's theory.\\
Suppose $\Phi\in D ({I\hspace {-1.5 mm}R}^n)$ with 
$D ({I\hspace {-1.5 mm}R}^n)$ the space of smooth(i.e.$C^\infty$) 
C-valued test functions on ${I\hspace {-1.5 mm}R}^n$ with compact 
support and
\begin{equation}
\label{math:2.6}
\int \Phi(x) dx =1.
\end{equation}
For $\epsilon>0$ we define the rescaled function $\Phi^\epsilon (x)$ as 
\begin{equation}
\label{math:2.7}
\Phi^\epsilon (x) = \frac{1}{\epsilon^n} \Phi (\frac{x}{\epsilon}).
\end{equation}
Now, for $f: {I\hspace {-1.5 mm}R}^n \longrightarrow
C$, not necessarily continuous, we define the smoothing process for 
$f$ as one of the convolutions
\begin{equation}
\label{math:2.8}
\tilde{f}(x) : = \int f(y) \Phi (y-x) d^n y,
\end{equation}
or
\begin{equation}
\label{math:2.9}
\tilde{f}_\epsilon (x) : = \int f(y) \Phi^\epsilon (y-x) d^n y.
\end{equation}
According to (7), equation (9) has the following explicit form
\begin{equation}
\label{math:2.10}
\tilde{f}_\epsilon (x) : = \int f(y) \frac{1}{\epsilon^n} \Phi (\frac{y-x}
{\epsilon}) d^n y.
\end{equation}
This smoothing procedure is valid for distributions too. Take the distribution 
$\,R\,$, then by  smoothing of $\,R\,$ we mean one of the two
convolutions (8) or (9) with $f$ replaced by$\,R\,$. 
Remember that $\,R\,$ is a C-valued functional such that
\begin{equation}
\label{math:2.11}
\Phi \in  D(I\hspace {-1.5 mm}R^n) \Longrightarrow (R, \Phi) \in C,
\end{equation}
where $(R,\Phi)$ is the convolution of $\,R\,$ and $\Phi$.\\
Now we can perform the product $\,Rf\,$ of the distribution 
$\,R\,$ with the discontinuous function $f$ through the action of the 
product on a test function $\,\Psi$. First we define the product of 
corresponding smoothed quantities $\tilde{R}_\epsilon$ with 
$\tilde{f}_\epsilon$ and then take the limit
\begin{equation}
\label{math:2.12}
(R f, \Psi) = \lim_{\epsilon \rightarrow 0}
\int \tilde{R}_\epsilon (x) \tilde{f}_\epsilon (x) \Psi (x) d^n x.
\end{equation}
The multiplication so defined does not coincide with the ordinary 
multiplication even for continuous functions. Colombeau's strategy 
to resolve this defficulty is as follows. Consider one-parameter 
families $(f_\epsilon)$ of $C^\infty$ functions  used to 
construct the  algebra
\begin{equation}
\label{math:2.13}
\begin{array}{ll}
\cal{E}_M\em(I\hspace {-1.5 mm}R^n) = &\{(f_\epsilon) \mid  
f_\epsilon \in
C^\infty ({I\hspace {-1.5 mm}R}^n) \quad \forall K \subset 
{I\hspace {-1.5 mm}R}^n \ compact, \\
&\forall \alpha \in {I\hspace {-1.5 mm}N}^n \quad \exists N \in
 {I\hspace {-1.5 mm}N}, \ \exists \eta >0, 
\ \exists c>0\\
&such \ that \ \sup_{x\in K} |D^\alpha f_\epsilon (x)| \leq c
 \epsilon^{-N} \quad 
\forall 0 < \epsilon < \eta \},
\end{array}
\end{equation}
where
\begin{equation}
\label{math:2.14}
D^\alpha = \frac{\partial^{|\alpha|}}{(\partial x^1)^{\alpha_1} \cdots 
(\partial x^n)^{\alpha_n}},
\end{equation}
and
$$
|\alpha| = \alpha_1 +  \alpha_2  + \cdots +\alpha_n.
$$
Accordingly, $C^\infty$-functions are embedded into $\cal{E}_M 
\em(I\hspace {-1.5 mm}R^n)$ as
constant sequences. For continuous functions and distributions we require a
smoothing kernel $\phi(x)$, such that
\begin{equation}
\label{math:2.15}
\int d^n x \ \varphi(x) \ dx = 1 \quad and\quad\int d^n x \ x^\alpha \varphi(x) 
= 0 \quad |\alpha|  \geq 1.
\end{equation}
Smoothing is defined as (10) for any  function $f$.\\
Now, we have to identify  different embeddings of $C^\infty$ functions. Take
a suitable ideal $\cal{N}$$({I\hspace {-1.5 mm}R}^n)$ defined as
\begin{equation}
\label{math:2.16}
\begin{array}{lll}
\cal{N} \em(I\hspace {-1.5 mm}R^n)
&=  \{(f_\epsilon)
 \mid (f_\epsilon) \in \cal{E}_M\em(I\hspace {-1.5 mm}R^n) \quad 
 \forall K \subset{I\hspace{-1.5 mm}R}^n \ compact, \\
&\forall \alpha \in {I\hspace {-1.5 mm}N}^n, \ \forall N \in   
\em I\hspace {-1.5 mm}N \quad \exists \ \eta>0 , \ \exists \ c>0,\\
&such \ that \ \sup_{x\in K}
|D^\alpha f_\epsilon (x)| \leq c \in^N \quad \forall 0<\epsilon<\eta \},
\end{array}
\end{equation} 
containing negligible functions such as 
\begin{equation}
\label{math:2.17}
f(x) - \int  d^n y \frac{1}{\epsilon^n} \varphi (\frac{y-x}{\epsilon}) f(y).
\end{equation}
Now, the Colombeau algebra $\cal{G}$$({I\hspace{-1.5 mm}R}^n)$ is defined as,
\begin{equation}
\label{math:2.18}
\cal{G}\em(I\hspace{-1.5 mm}R^n) = \frac{\cal{E}_M \em({I\hspace{-1.5 mm}R}^n)}
{\cal{N}\em({I\hspace{-1.5 mm}R}^n)}
\end{equation}
A Colombeau generalized function is thus a moderate family $(f_\epsilon (x))$
of $C^\infty$ functions modulo negligible families. Two Colombeau objects 
$(f_\epsilon)$ and $(g_\epsilon)$ are said to be associate 
(written as $(g_\epsilon) \approx (f_\epsilon)$) if
\begin{equation}
\label{math:2.19}
\begin{array}{ll}
\lim_{\epsilon \rightarrow 0} \ \int d^n x \ (f_\epsilon (x) -  g_\epsilon (x)) 
\ \varphi (x)  = 0
\quad \forall \ \varphi \in D (I\hspace {-1.5 mm}R^n).
\end{array}
\end{equation}
For example, if $\varphi(x) = \varphi{(-x)}$ then $\delta \theta \approx
\frac{1}{2} \delta$, where $\delta$ is Dirac delta function and $\theta$ is
Heaviside Step function. Moreover, we have in this algebra 
$\theta^n \approx \theta$ and not $\theta^n = \theta$. For an extensive  
introduction to Colombeau theory, see[24,25].
\newpage

\section{Distributional Approach to Signature Change}
There are two methods of handling singular hypersufaces in general 
relativity. The mostly used method of Darmois-Israel, based on the 
Gauss-Kodazzi decomposition of space-time, is handicapped through the
junction conditions which make the formalism unhandy. For our purposes
the distributional approach of Mansouri and Khorrami (M-Kh) [23] 
is the most suitable one. In this formalism  the whole spacetime,
including the singular hypersurface, is treated with a unified metric 
without bothering about the junction conditions along the hypersurface. 
These conditions are shown to be authomatically fulfilled as part of the 
field equations. 
In the M-Kh-distributional approach one choose special coordinates 
which are continuous along the singular hypersuface to avoid non-linear 
operations of distributiuons. Here, using Colombeau algebra, which 
allows for non-linear operations of distributions, we
generalize the M-Kh method to the special case of signature changing
cosmological models.\\
Consider a spacetime with the following FRW metric containing 
a steplike lapse function
\begin{equation}
\label{math:3.1}
ds^2 = -f(t) dt^2 + R^2(t) (\frac{dr^2}{1-kr^2} + r^2 d \theta^2 +  
r^2 \sin^2 \theta d \varphi^2),
\end{equation}
where
\begin{equation}
\label{math:3.2}
f(t) =  \theta (t) - \theta (-t).  
\end{equation}
It describes a signature changing spacetime with the singular surface 
$t=0.$ The metric describes a Riemanian space for $t<0$ and a Lorenzian
space-time for $t>0$. As has been argued before $f(t)$, analogous to 
$\theta(t)$, has a microscopic structure around $t=0$ with a nonvanishing 
jump, $\tau$ . We choose again 
\begin{equation}
\label{math:3.4}
\theta(t)\mid_{t=0} = \epsilon   \quad \quad with \quad  \epsilon > \frac{1}{2}. 
\end{equation}
Since $\theta(-t) = 1 - \theta(t) $, we have $\theta(-t) = 1 - \epsilon $  
and
\begin{equation}
\label{math:3.5}
f(t) = 2 \epsilon - 1 .
\end{equation}
This value gives us the correct change of sign in going from $t<0$ 
to $t>0$. This ``regularization'' of $f(t)$ at $t=0$ allows us to use 
operations such as $f(t)^{-1}$, $f(t)^{2}$, and $|f(t)|^{-1}$. The physical 
interpretation of this behaviour of $\,f(t)\,$ is that the 
phenomenon of signature change occurs as a quantum mechanical tunneling 
effect. This tunneling occurs in a width equal to the width of 
the jump, i.e. $\tau$.\\ 
In what follows we consider $\,f(t)\,$ to be the regularized function 
$\tilde{f}_\epsilon$, defined according to Colombeau's algebra.
Now, we are prepared to calculate the dynamics of the signature changing  
hypersurface in the line of M-Kh procedure[23]. First we calculte the 
relevant components of the Einstein tensor, $G_{tt}$  
and $G_{rr}$, for the metric (20):
$$G_{tt} = -\frac{3}{2} \ \frac{-2\ddot{R} f + \dot{f} \dot{R}}{fR}  +
\frac{1}{2} f \Bigl\{ \frac{3}{2} \frac{-2\ddot{R}f + \dot{f}\dot{R}}{f^2 R}$$
\begin{equation}
\label{math:3.4}
-\frac{3}{2} \frac{2Rf\ddot{R}-R\dot{f}\dot{R} + 4(kf^2 + f \dot{R}^2)}
{R^2 f^2}\Bigr\},
\end{equation}
and
$$G_{rr} = -\frac{1}{2} \ \frac{2 R\ddot{R} f - R\dot{f}\dot{R} +
4f\dot{R}^2 + 4kf^2}{f^2 (1-kr^2)}$$                           
\begin{equation}
\label{math:3.5}
-\frac{1}{2} \frac{R^2}{1-kr^2} 
\Bigl\{\frac{3}{2} \frac{-2\ddot{R}f + \dot{R}\dot{f}}{f^2 R}
-\frac{3}{2} \frac{2R\ddot{R}f-R\dot{f}\dot{R} + 4(kf^2 + f \dot{R}^2)}
{R^2 f^2}\Bigr\}
\end{equation}
According to standard calculus of distributions, we have
\begin{equation}
\label{math:3.6}
\begin{array}{ll}
&\dot{f}(t)\ = \dot{\theta}(t) - \dot{\theta}(-t)\\
&\quad\, \,\, \ = \delta(t) + \delta(-t) = 2 \ \delta(t),
\end{array}
\end{equation}
taking into account $\,\delta(-t)=\delta(t)\,$. Now, using Colombeau 
algebra we can write 
\begin{equation}
\label{math:3.7}
\theta (t) \delta (t) \approx \frac{1}{2} \delta (t).
\end{equation}
Therefore we may write
\begin{equation}
\begin{array}{ll}
\label{math:3.8}
f(t) \ \delta (t) &= \theta(t) \delta(t) - \theta(-t) \delta(t)\\
&\approx \frac{1}{2} \delta(t) - \frac{1}{2} \delta(t)\\
&\approx 0,
\end{array}
\end{equation}
In evaluating (24, 25) we should take care of the following property of
association. Having
$$AB\approx AC,$$ we are not allowed to conclud $$B\approx C.$$ Now, using  
the relations (27) and (28) we obtain for the singular parts of equations 
(24) and (25) 
\begin{equation}
\label{math:3.9}
\hat{G}_{tt} = 0,
\end{equation}             
\begin{equation}
\label{math:3.10}
\hat{G}_{rr} = -\frac{2 R\dot{R}}{f^2 (1-k r^2)} \ \delta(t).
\end{equation}
where multiplication of the distribution $\delta(t)$ with the discontinuous 
functions $\frac{1}{f^2}$ is defined as in (12).

According to [23] the complete energy-momentum tensor can be written as
\begin{equation}
\label{math:3.11}
T_{\mu\nu} = \theta(t) \ T^+_{\mu\nu} \ + \ \theta(-t) \ T^-_{\mu\nu} + \ C 
S_{\mu\nu} \ \delta(t),
\end{equation}
where $T_{\mu\nu}^{\pm}$ are energy-momentum tensors corresponding to Riemannian 
and Lorentzian regions respectively and $C$ is a constant which can be obtained 
by taking the following pill-box integration defining $S_{\mu\nu}$[23]:
\begin{equation}
\label{math:3.12}
S_{\mu\nu} = \lim_{\Sigma \rightarrow 0} \int_{-\Sigma}^{\Sigma}(T_{\mu\nu}
- g_{\mu\nu}\frac{\Lambda}{\kappa})dn = \frac{1}{\kappa}\lim_{\Sigma \rightarrow 0}
\int_{-\Sigma}^{\Sigma} G_{\mu\nu} dn,
\end{equation}
Since
\begin{equation}
\label{math:3.13}
\hat{T}_{\mu\nu} = CS_{\mu\nu}\delta(\Phi(x)),
\end{equation}
and
\begin{equation}
\label{math:3.14}
\int{\hat{T}_{\mu\nu}} dn = CS_{\mu\nu}\int{\delta(\Phi(x))}dn = 
CS_{\mu\nu}|\frac{dn}{d\Phi}|,
\end{equation}
we find 
\begin{equation}
\label{math:3.12}
C = | \frac{d\Phi}{dn} | = |n^\mu \ \partial_\mu \ \Phi|,
\end{equation}
where $\Phi = t = 0$ defines the singular surface $\Sigma$. The vector $n_\mu$ 
is normal to the surface $\Phi$ and $n$ measure the distance along it.
Using the metric (20) we obtain
\begin{equation}
\label{math:3.13}
C = \frac{1}{|f(t)|}. 
\end{equation}
The distributional part of the Einstein equation reads as follows:
\begin{equation}
\label{math:3.15}
\hat{G}_{\mu\nu} =  \kappa \hat{T}_{\mu\nu}.
\end{equation}
we obtain using equations (29,30,33,36,37):
\begin{equation}
\label{math:3.16}
0 = \frac{\kappa}{|f(t)|} \ S_{tt} \delta (t),
\end{equation}
and
\begin{equation}
\label{math:3.17}
-\frac{2R\dot{R}}{f^2 (1- k r^2)} \ \delta (t) = \frac{\kappa}{|f(t)|} \ S_{rr} 
\ \delta (t).
\end{equation}
Now using equation (12), we must define the multiplication of $\delta$-distribution 
with the discontinuous functions $\frac{1}{|f|}$ and $\frac{1}{f^2}$. To this end 
we consider them as Colombeau's regularized functions,
\begin{equation}
\label{math:3.18}
\tilde{G}_{1\epsilon}(t) = {\delta}_{\epsilon}(t) {(\frac{1}
{|f(t)|})}_{\epsilon}
\end{equation}
and
\begin{equation}
\label{math:3.19}
\tilde{G}_{2\epsilon}(t) = {\delta}_{\epsilon}(t) {(\frac{1}
{f^{2}(t)})}_{\epsilon}.    
\end{equation}
Now according to (12), these two multiplications are as follows,
\begin{equation}
\label{math:3.20}
(\delta(t)\frac{1}{|f(t)|}, \Psi) = \lim_{\epsilon \rightarrow 0}
\int \tilde{G}_{1\epsilon (t)} \Psi (t) dt
\end{equation}
and
\begin{equation}
\label{math:3.21}
(\delta(t)\frac{1}{f^{2}(t)}, \Psi) = \lim_{\epsilon \rightarrow 0}
\int \tilde{G}_{2\epsilon (t)} \Psi (t) dt
\end{equation}
for any test function, $\Psi$.
Now we argue that $\tilde{G}_{1\epsilon}$ and $\tilde{G}_{2\epsilon}$ are   
associate in Colombeau's sense, i.e.
\begin{equation}
\label{math:3.22}
\lim_{\epsilon \rightarrow 0}\int (\tilde{G}_{1\epsilon}(t) -
\tilde{G}_{2\epsilon}(t) )\Psi (t) dt = 0.
\end{equation}
This is correct because although $\tilde{G}_{1\epsilon}$ and $\tilde{G}_{2\epsilon}$ 
are divergent at a common point, the difference in their `` microscopic 
structure'' at that point tends to zero for $\epsilon \rightarrow 0$. 
Therefore, we obtain from (38), (39), and (44) the final form of the 
energy-momentum tensor of the singular surface, or the dynamics of $\Sigma$:
\begin{equation}
\label{math:3.23}
S_{tt} = 0,
\end{equation}
and
\begin{equation}
\label{math:3.24}
S_{rr} = -\frac{2R\dot{R}}{\kappa (1-k r^2)}.
\end{equation}
We therefore see that the energy-momentum tensor of the singular hypersurface
is not vanishing in general.

\section{Concluding Remarks}
We are now in a position to resolve the controversy between Hayward and 
Hellaby $\&$ Dray[14,18,19]. In general, we have seen that the energy-momentum
tensor of the hypersurface does not vanish and is proportional to $\dot R$.
Therefore, if $\dot R = 0$ then $S_{\mu \nu} = 0$, but there is no reason 
to assume $\dot R = 0$. This is in accordance with the results of 
Hellaby $\&$ Dray.

\end{document}